\documentclass[12pt]{article}%
\usepackage{amsfonts}
\usepackage{amsmath}
\usepackage{geometry}
\usepackage{amssymb}
\usepackage{graphicx}
\usepackage[onehalfspacing]{setspace}
\usepackage{hyperref}
\usepackage[nosort]{cite}%
\providecommand{\U}[1]{\protect\rule{.1in}{.1in}}

\begin{document}

\title{\textbf{Smarr's Formula in Eleven-Dimensional Supergravity\medskip}}
\author{\textbf{Patrick A. Haas\medskip}\\Department of Physics and Astronomy\\University of Southern California\\Los Angeles, CA 90089, USA\\\medskip\\phaas@usc.edu}
\date{}
\maketitle

\begin{abstract}
\noindent\thispagestyle{empty}We examine the Smarr formula in
eleven-dimensional spacetime compactified on a general six-dimensional,
Ricci-flat manifold. We show that non-zero mass for smooth and horizonless
solutions can only be provided by cohomology. Furthermore, we confirm the
result that there are no solitons without topology and prove the fact that
Chern-Simons terms in the mass formula only appear in order to generate a
purely topological integral.\newpage\setcounter{page}{1}

\end{abstract}
\tableofcontents

\section{Introduction}

In quantum mechanics the main issues associated with the classical description
of a black hole are the singularity and the \textquotedblleft information
paradox\textquotedblright\ following from the problem of accounting for the
entropy stored inside and associated with the horizon. One way physicists have
addressed this is by searching for a consistent quantum description of a black hole.

In 2002, Samir Mathur proposed a concept within the framework of string
theory: The \textquotedblleft fuzzball\textquotedblright\ program
\cite{Mathur}, which leads in the supergravity limit to smooth, horizonless,
and asymptotically flat solutions representing solitons if time-independent.

Supergravity solitons have long been doubted to exist, especially since some
theorems appear to rule them out. In \cite{Gibbons:2013tqa} and \cite{Kunduri}%
\ it was recently shown that these theorems can be circumvented by allowing
non-trivial topology on the solution's spatial hypersurfaces. In addition to
topology the Chern-Simons interactions were shown to generally contribute to
solitonic solutions. In the current paper we examine these matters in an
eleven-dimensional spacetime, six dimensions of which are compactified. We
also show that the Chern-Simons terms only play a secondary role in combining
all the fluxes in the integral of the mass formula to a closed differential
form and thus ensure the soliton's mass to be purely topological.

Hence, for the existence of solitons in supergravity one still has
\textquotedblleft No solitons without topology\textquotedblright, and the
Chern-Simons interactions entirely support this circumstance without adding
any extra support mechanisms.

In Section 2 we write out the bosonic part of the eleven-dimensional
supergravity action \cite{Sugra11D} allowing an arbitrary constant in front of
the Chern-Simons term, and reexamine the work in \cite{Gibbons:2013tqa} under
these more general circumstances.

In Section 3 we set up the eleven-dimensional Komar integral, determine its
normalization based on the known result in five-dimensions
\cite{Gibbons:2013tqa}, and derive the eleven-dimensional version of Smarr's
formula reflecting the mass of the solitonic solutions purely by topology.

In Section 4 we recap the five-dimensional calculations from
\cite{Gibbons:2013tqa}, also allowing for non-trivial 1-forms and an arbitrary
Chern-Simons coefficient, in order to show both the purely topological nature
of the mass and the process of getting there from the more general
eleven-dimensional case by special choices of fields and compact
geometry\footnote{For detailed elaboration on flux compactification, see
\cite{Grana, Samtleben, Denef}.}.

\section{Preliminaries}

\subsection{The eleven-dimensional Supergravity action and equations of
motion}

Eleven-dimensional supergravity has in its bosonic sector the graviton,
$g_{MN}$, and the 3-form potential, $C_{MNK}=C_{\left[  MNK\right]  }%
$.\footnote{We indicate the whole eleven-dimensional spacetime with a capital
latin, the non-compact five-dimensional spacetime with a greek, and the
compact six-dimensional space with a small latin index.} The latter gives rise
to a field strengh 4-form, $F=dC$.

The bosonic action \cite{Sugra11D} is%
\begin{equation}
S_{11}=\int d^{11}x\sqrt{-g}\left(  R-\tfrac{1}{2}\left\vert F\right\vert
^{2}\right)  -\tfrac{\alpha}{6}\int C\wedge F\wedge F, \label{11D-action}%
\end{equation}
where we have introduced an arbitrary constant coefficient, $\alpha$, for the
Chern-Simons term. Supersymmetry corresponds to $\alpha=1$ but we wish to
examine the impact of the Cherm-Simons interactions in a broader,
non-supersymmetric version of the theory.

The Einstein equations resulting from $\left(  \ref{11D-action}\right)  $ are%
\begin{equation}
R_{MN}-\tfrac{1}{2}g_{MN}R=\tfrac{1}{12}F_{MRSK}F_{N}^{\text{ \ }RSK}%
-\tfrac{1}{96}g_{MN}F_{RSKL}F^{RSKL}, \label{Einstein}%
\end{equation}
which may be written,%
\begin{equation}
R_{MN}=\tfrac{1}{12}F_{MRSK}F_{N}^{\text{ \ }RSK}-\tfrac{1}{144}g_{MN}%
F_{RSKL}F^{RSKL}. \label{Einstein_rearranged}%
\end{equation}

The Maxwell equations resulting from $\left(  \ref{11D-action}\right)  $ are%
\begin{equation}
\nabla_{N}F_{\text{ }RSK}^{N}=J_{RSK}^{CS,\alpha}, \label{Maxwell}%
\end{equation}
with the eleven-dimensional Chern-Simons 3-form current\footnote{The
Levi-Civita tensor for curved spacetimes is related to the Levi-Civita symbol
of Minkowski spacetime like $\bar{\epsilon}^{M_{1}...M_{11}}=\left(
-g\right)  ^{-\frac{1}{2}}\epsilon^{M_{1}...M_{11}}\Leftrightarrow
\bar{\epsilon}_{M_{1}...M_{11}}=\left(  -g\right)  ^{\frac{1}{2}}%
\epsilon_{M_{1}...M_{11}}$.},%
\begin{equation}
J_{RSK}^{CS,\alpha}=\tfrac{\alpha}{1152}\bar{\epsilon}_{RSKM_{4}...M_{11}%
}F^{M_{4}...M_{7}}F^{M_{8}...M_{11}}. \label{CS-current}%
\end{equation}

Define a dual 7-form,%
\begin{equation}
G=\star_{11}F\Leftrightarrow G_{M_{1}...M_{7}}=\tfrac{1}{24}\bar{\epsilon
}_{M_{1}...M_{11}}F^{M_{8}...M_{11}}. \label{G}%
\end{equation}
The equation of motion for $G$ is simply the Bianchi identity for $F$ and vice
versa,%
\begin{equation}
\nabla_{R}G^{M_{1}...M_{6}R}=\tfrac{1}{24}\bar{\epsilon}^{M_{1}...M_{6}%
RM_{8}...M_{11}}\nabla_{R}F_{M_{8}...M_{11}}=0,
\end{equation}
and, with $\left(  \ref{Maxwell}\right)  $,%
\begin{equation}
\nabla_{\lbrack M_{4}}G_{M_{5}...M_{11}]}=\tfrac{35\alpha}{8}F_{[M_{4}%
...M_{7}}F_{M_{8}...M_{11}]}\Leftrightarrow dG=\tfrac{\alpha}{2}F\wedge F.
\label{dG}%
\end{equation}

Note that%
\begin{equation}
G_{MS_{1}...S_{6}}G^{NS_{1}...S_{6}}=-150\delta_{\lbrack M}^{N}F_{K_{1}%
...K_{4}]}F^{K_{1}...K_{4}}=120F_{MK_{2}K_{3}K_{4}}F^{NK_{2}K_{3}K_{4}%
}-30\delta_{M}^{N}F_{K_{1}...K_{4}}F^{K_{1}...K_{4}},
\end{equation}
which allows us to rewrite $\left(  \ref{Einstein_rearranged}\right)  $ as%
\begin{equation}
R_{MN}=\tfrac{1}{18}F_{MRSK}F_{N}^{\text{ }RSK}+\tfrac{1}{4320}G_{MS_{1}%
...S_{6}}G_{N}^{\text{ }S_{1}...S_{6}}. \label{Einstein_rearranged_2}%
\end{equation}

\subsection{Invariances}

We assume that the matter fields have the symmetries of the metric, that is,
they are invariant under a Killing vector, $K$,%
\begin{equation}
\mathcal{L}_{K}F=0=\mathcal{L}_{K}G, \label{Invariance}%
\end{equation}
where $\mathcal{L}_{K}$ is the corresponding Lie derivative. Cartan's magic
formula,%
\begin{equation}
\mathcal{L}_{K}\omega=d\left(  i_{K}\omega\right)  +i_{K}\left(
d\omega\right)  , \label{Lie derivative}%
\end{equation}
applied to the 4-form $F$, yields%
\begin{equation}
0=d\left(  i_{K}F\right)  \Leftrightarrow K^{M}F_{MNRS}=3\nabla_{\lbrack
N}\lambda_{RS]}+H_{NRS}^{\left(  3\right)  }, \label{F_Killing}%
\end{equation}
where $\lambda$ are the magnetostatic 2-form potentials of $G$ and the
electrostatic 2-form potentials of $F$, respectively, and $H^{\left(
3\right)  }$ is a closed but not exact 3-form, that is, $H^{\left(  3\right)
}\in H^{3}\left(  \mathcal{M}_{11}\right)  $.

For $G$ we find%
\begin{align}
0  &  =d\left(  i_{K}G\right)  +i_{K}\left(  dG\right)  \Leftrightarrow
d\left(  i_{K}G\right)  =-\alpha\left(  d\lambda+H^{\left(  3\right)
}\right)  \wedge F=-\alpha d\left(  \lambda\wedge F-H^{\left(  3\right)
}\wedge C\right) \nonumber\\
&  \Leftrightarrow d\left(  i_{K}G+\alpha\lambda\wedge F-\alpha H^{\left(
3\right)  }\wedge C\right)  =0,
\end{align}
and so%
\begin{equation}
K^{M}G_{MM_{1}...M_{6}}=6\nabla_{\lbrack M_{1}}\Lambda_{M_{2}...M_{6}%
]}-15\alpha\lambda_{\lbrack M_{1}M_{2}}F_{M_{3}...M_{6}]}+20\alpha
H_{[M_{1}M_{2}M_{3}}^{\left(  3\right)  }C_{M_{4}M_{5}M_{6}]}+H_{M_{1}%
...M_{6}}^{\left(  6\right)  }, \label{G_Killing}%
\end{equation}
where $\Lambda$ is a generic 5-form and $H^{\left(  6\right)  }\in
H^{6}\left(  \mathcal{M}_{11}\right)  $ a closed but not exact 6-form.

From $\left(  \ref{F_Killing}\right)  $ and $\left(  \ref{G_Killing}\right)  $
follows%
\begin{align}
K^{M}F_{MRSK}F_{N}^{\text{ \ }RSK}  &  =-3\nabla_{R}\left(  \lambda
^{SK}F_{\text{ }SKN}^{R}\right)  +\tfrac{\alpha}{384}\lambda^{SK}\bar
{\epsilon}_{SKNM_{4}...M_{11}}F^{M_{4}...M_{7}}F^{M_{8}...M_{11}}\nonumber\\
&  +H_{RSK}^{\left(  3\right)  }F_{N}^{\text{ \ }RSK}\\
K^{M}G_{MS_{1}...S_{6}}G_{N}^{\text{ \ }S_{1}...S_{6}}  &  =6\nabla_{S_{1}%
}\left(  \Lambda_{S_{2}...S_{6}}G_{N}^{\text{ \ }S_{1}...S_{6}}\right)
-\tfrac{15\alpha}{24}\bar{\epsilon}_{NS_{1}...S_{10}}\lambda^{S_{1}S_{2}%
}F^{S_{3}...S_{6}}F^{S_{7}...S_{10}}\nonumber\\
&  +\left(  20\alpha H_{S_{1}S_{2}S_{3}}^{\left(  3\right)  }C_{S_{4}%
S_{5}S_{6}}+H_{S_{1}...S_{6}}^{\left(  6\right)  }\right)  G_{N}^{\text{
\ }S_{1}...S_{6}},
\end{align}
and hence, the Einstein equations $\left(  \ref{Einstein_rearranged_2}\right)
$ become%
\begin{align}
K^{M}R_{MN}  &  =-\tfrac{1}{720}\nabla_{R}\left(  120\lambda^{SK}F_{\text{
}SKN}^{R}-\Lambda^{S_{2}...S_{6}}G_{\text{ }S_{2}...S_{6}N}^{R}\right)
+\tfrac{1}{18}H_{RSK}^{\left(  3\right)  }F_{N}^{\text{ \ }RSK}\nonumber\\
&  +\tfrac{1}{4320}\left(  20\alpha H_{S_{1}S_{2}S_{3}}^{\left(  3\right)
}C_{S_{4}S_{5}S_{6}}+H_{S_{1}...S_{6}}^{\left(  6\right)  }\right)
G_{N}^{\text{ \ }S_{1}...S_{6}}. \label{Ricci_Killing}%
\end{align}
As in \cite{Gibbons:2013tqa} the $\lambda\left(  \star F\wedge F\right)  $
terms cancel out, and it is important to note that this happens independently
of the choice of the parameter, $\alpha$. However, explicit Chern-Simons terms
indeed go along with the inclusion of $H^{\left(  3\right)  }$. As we will
describe below, the analogue of this in the analysis of \cite{Gibbons:2013tqa}
was omitted for the assumption of simple-connectedness of the four-dimensional
slices, $\Sigma$.

\section{Komar integrals in eleven-dimensional supergravity}

If $K$ is a Killing vector, then the Komar integral
\cite{Gibbons:1993xt,Sabra:1997yd,Myers:1986un,Peet:2000hn,Gibbons:2013tqa} is%
\begin{equation}
\int_{\partial\Sigma}\star dK=\int_{\partial\Sigma}\left(  \partial_{M}%
K_{N}-\partial_{N}K_{M}\right)  d\Sigma^{MN}=-2\int_{\Sigma}R_{MN}K^{M}%
d\Sigma^{N}. \label{Komar integral}%
\end{equation}
If $K$ is timelike at infinity, we can use a coordinate with $K\approx
\tfrac{\partial}{\partial t}$, so near infinity the 1-form is then%
\begin{equation}
K\approx g_{00}dt. \label{Killing vector_infinity}%
\end{equation}

We know how to get the conserved mass, $M$, from the five-dimensional Komar
integral in \cite{Gibbons:2013tqa}, assuming%
\begin{equation}
M=\int_{\Sigma_{D-1}}T_{00}^{\left(  D\right)  }d\Sigma^{\left(  D-1\right)
}=A_{D}\int_{S^{D-2}}\star_{D}dK,
\end{equation}
where $A_{D}$ is a normalization, and in particular,%
\begin{equation}
A_{5}=-\tfrac{3}{32\pi G_{5}}.
\end{equation}

We can set up the formula for eleven dimensions,%
\begin{equation}
M=A_{11}\int_{S^{9}}\star_{11}dK=A_{11}\int_{S^{3}\times M_{6}}\star
_{5}dK\wedge dvol_{6}|_{r=\infty}=A_{11}vol_{6}\int_{S^{3}}\star_{5}dK,
\end{equation}
and conclude the relation of the normalization factors directly,%
\begin{equation}
A_{11}=\tfrac{A_{5}}{vol_{6}}=-\tfrac{3}{32\pi G_{5}vol_{6}},
\end{equation}
where $vol_{6}$ is the volume of the $M_{6}$ at space's infinity.

Hence, the eleven-dimensional Komar integral is:%
\begin{equation}
M=-\tfrac{3}{32\pi G_{5}vol_{6}}\int_{S^{9}}\star_{11}dK=\tfrac{3}{16\pi
G_{5}vol_{6}}\int_{\Sigma_{10}}R_{MN}K^{M}d\Sigma^{N}.
\label{Komar integral_normalized}%
\end{equation}
Using $\left(  \ref{Ricci_Killing}\right)  $ in $\left(
\ref{Komar integral_normalized}\right)  $ and assuming that the boundary terms
fall off sufficiently fast at infinity, the conserved mass is given by%
\begin{equation}
M=\tfrac{3}{16\pi G_{5}vol_{6}}\int_{\Sigma_{10}}\left[  \tfrac{1}{18}%
H_{RSK}^{\left(  3\right)  }F_{N}^{\text{ \ }RSK}+\tfrac{1}{4320}\left(
20\alpha H_{S_{1}S_{2}S_{3}}^{\left(  3\right)  }C_{S_{4}S_{5}S_{6}}%
+H_{S_{1}...S_{6}}^{\left(  6\right)  }\right)  G_{N}^{\text{ \ }S_{1}%
...S_{6}}\right]  d\Sigma^{N}. \label{Komar mass_no interior boundaries}%
\end{equation}

Rewritten in terms of differential forms, this becomes%
\begin{equation}
M=\tfrac{1}{32\pi G_{5}vol_{6}}\int_{\Sigma_{10}}\left[  H^{\left(  3\right)
}\wedge\left(  2G-\alpha C\wedge F\right)  -H^{\left(  6\right)  }\wedge
F\right]  . \label{Komar mass_topological}%
\end{equation}
Note that the differential form,%
\[
2G-\alpha C\wedge F,
\]
is closed by virtue of $\left(  \ref{dG}\right)  $. Since $F$, $H^{\left(
3\right)  }$ and $H^{\left(  6\right)  }$ are closed by definition, the
integral is purely topological for all values of $\alpha$. This means that the
contribution of explicit Chern-Simons terms in the Komar mass formula is
precisely to turn the latter into an integral over cohomology.

\section{Recap of the five-dimensional case}

We are going to repeat the calculations done in \cite{Gibbons:2013tqa} and in
addition allow for non-trivial 1-forms and an arbitrary constant coefficient
of the Chern-Simons term. Finally, we will compare this to the
eleven-dimensional case.

The action \cite{Bena&Warner:2007, Niehoff:2013mla} is%
\begin{equation}
S=\int\left(  \star_{5}\mathcal{R}-Q_{IJ}dX^{I}\wedge\star_{5}dX^{J}%
-Q_{IJ}\mathcal{F}^{I}\wedge\star_{5}\mathcal{F}^{J}-\tfrac{1}{6}%
C_{IJK}\mathcal{F}^{I}\wedge\mathcal{F}^{J}\wedge A^{K}\right)  ,
\end{equation}
where $C_{IJK}=\left\vert \epsilon_{IJK}\right\vert $ and $X^{I}$, $I=1,2,3$,
are scalar fields arising from reducing the eleven-dimensional metric,%
\begin{equation}
ds_{11}^{2}=ds_{5}^{2}+\left(  \tfrac{Z_{2}Z_{3}}{Z_{1}^{2}}\right)
^{\frac{1}{3}}\left(  dx_{5}^{2}+dx_{6}^{2}\right)  +\left(  \tfrac{Z_{1}%
Z_{3}}{Z_{2}^{2}}\right)  ^{\frac{1}{3}}\left(  dx_{7}^{2}+dx_{8}^{2}\right)
+\left(  \tfrac{Z_{1}Z_{2}}{Z_{3}^{2}}\right)  ^{\frac{1}{3}}\left(
dx_{9}^{2}+dx_{10}^{2}\right)  , \label{Metric_11D}%
\end{equation}
with the reparametrization,%
\begin{equation}
X^{1}=\left(  \tfrac{Z_{2}Z_{3}}{Z_{1}^{2}}\right)  ^{\frac{1}{3}},\text{
}X^{2}=\left(  \tfrac{Z_{1}Z_{3}}{Z_{2}^{2}}\right)  ^{\frac{1}{3}},\text{
}X^{3}=\left(  \tfrac{Z_{1}Z_{2}}{Z_{3}^{2}}\right)  ^{\frac{1}{3}},
\label{Reparameterization}%
\end{equation}
to fulfill the constraint $X^{1}X^{2}X^{3}=1.$

Moreover, there is a metric for the kinetic terms,%
\begin{equation}
Q_{IJ}=\tfrac{1}{2}\text{diag}\left(  \left(  \tfrac{1}{X^{1}}\right)
^{2},\left(  \tfrac{1}{X^{2}}\right)  ^{2},\left(  \tfrac{1}{X^{3}}\right)
^{2}\right)  , \label{Metric_kinetic terms}%
\end{equation}
necessary also for the dualization of the field strength\footnote{We use
calligraphic script at some places to avoid confusion between the five- and
eleven-dimensional objects.}, $\mathcal{F}^{I}=dA^{I}$,%
\begin{equation}
\mathcal{G}_{I}=Q_{IJ}\left(  \star_{5}\mathcal{F}^{J}\right)  . \label{G_5D}%
\end{equation}
The analysis done for equation (5.8) in \cite{Gibbons:2013tqa} leads with an
arbitrary Chern-Simons coefficient to%
\begin{equation}
d\mathcal{G}_{I}=\tfrac{\beta}{4}C_{IJK}\mathcal{F}^{J}\wedge\mathcal{F}^{K}.
\label{dG_5D}%
\end{equation}

\subsection{Incorporating 1-forms}

In \cite{Gibbons:2013tqa} non-trivial 1-forms have not been considered, since
their contribution was assumed to not provide new interesting physics, but
here we incorporate them for completeness of the further below stated
dictionary of the fields. In particular, they correspond to the
eleven-dimensional 3-form.

Equation (5.13) of \cite{Gibbons:2013tqa} can be extended to%
\begin{equation}
K^{\rho}\mathcal{F}_{\rho\mu}^{I}=\partial_{\mu}\lambda^{I}+H_{\mu}^{\left(
1\right)  I}. \label{F_Killing_5D}%
\end{equation}
As a consequence we get%
\begin{align}
d\left(  i_{K}\mathcal{G}_{I}\right)   &  =-i_{K}d\mathcal{G}_{I}%
=-\tfrac{\beta}{4}C_{ILM}i_{K}\left(  \mathcal{F}^{L}\wedge\mathcal{F}%
^{M}\right)  =-\tfrac{\beta}{2}C_{ILM}\left(  d\lambda^{L}+H^{\left(
1\right)  L}\right)  \wedge\mathcal{F}^{M}\nonumber\\
&  =-\tfrac{\beta}{2}C_{ILM}d\left(  \lambda^{L}\mathcal{F}^{M}-H^{\left(
1\right)  L}\wedge A^{M}\right)  ,
\end{align}
so%
\begin{equation}
K^{\rho}\mathcal{G}_{I\rho\mu\nu}=2\partial_{\lbrack\mu}\Lambda_{I\nu]}%
-\tfrac{\beta}{2}C_{ILM}\left(  \lambda^{L}\mathcal{F}_{\mu\nu}^{M}-2H_{[\mu
}^{\left(  1\right)  L}A_{\nu]}^{M}\right)  +H_{I\mu\nu}^{\left(  2\right)  }.
\label{G_Killing_5D}%
\end{equation}
Note, that also here we included an arbitrary constant coefficient, $\beta$,
in front of the Chern-Simons term for which, like in the eleven-dimensional
case, $\beta=1$ means supersymmetry.

It follows%
\begin{align}
K^{\mu}\left(  Q_{IJ}\mathcal{F}_{\mu\rho}^{I}\mathcal{F}_{\nu}^{J\text{ }%
\rho}\right)   &  =\nabla_{\rho}\left(  Q_{IJ}\lambda^{I}\mathcal{F}_{\nu
}^{J\text{ }\rho}\right)  +\tfrac{\beta}{16}C_{IJK}\bar{\epsilon}_{\nu
\alpha\beta\gamma\delta}\lambda^{I}\mathcal{F}^{J\alpha\beta}\mathcal{F}%
^{K\gamma\delta}+Q_{IJ}H_{\rho}^{\left(  1\right)  I}\mathcal{F}_{\nu
}^{J\text{ }\rho}\\
K^{\mu}\left(  Q^{IJ}\mathcal{G}_{I\mu\rho\sigma}\mathcal{G}_{J}^{\nu
\rho\sigma}\right)   &  =2\nabla_{\rho}\left(  Q^{IJ}\Lambda_{I\sigma
}\mathcal{G}_{J}^{\nu\rho\sigma}\right)  -\tfrac{\beta}{4}C_{ILM}\bar
{\epsilon}^{\nu\alpha\beta\rho\sigma}\lambda^{I}\mathcal{F}_{\alpha\beta}%
^{L}\mathcal{F}_{\rho\sigma}^{M}\nonumber\\
&  +Q^{IJ}\left(  \beta C_{ILM}H_{\rho}^{\left(  1\right)  L}A_{\sigma}%
^{M}+H_{I\rho\sigma}^{\left(  2\right)  }\right)  \mathcal{G}_{J}^{\nu
\rho\sigma},
\end{align}
and hence%
\begin{align}
K^{\mu}R_{\mu\nu}  &  =\tfrac{1}{3}\nabla_{\rho}\left(  2Q_{IJ}\lambda
^{I}\mathcal{F}_{\nu}^{J\text{ }\rho}+Q^{IJ}\Lambda_{I\sigma}\mathcal{G}%
_{J\nu}^{\text{ \ \ }\rho\sigma}\right)  +\tfrac{2}{3}Q_{IJ}H^{\left(
1\right)  I\rho}\mathcal{F}_{\nu\rho}^{J}\nonumber\\
&  +\tfrac{1}{6}Q^{IJ}\left(  \beta C_{ILM}H^{\left(  1\right)  L\rho
}A^{M\sigma}+H_{I}^{\left(  2\right)  \rho\sigma}\right)  \mathcal{G}%
_{J\nu\rho\sigma}.
\end{align}
Excluding inner boundaries, the Komar mass formula becomes%
\begin{equation}
\tfrac{16\pi G_{5}}{3}M=\int_{\Sigma_{4}}\mathcal{R}_{\mu\nu}K^{\mu}%
d\Sigma^{\nu}=\int_{\Sigma_{4}}\left[  \tfrac{2}{3}Q_{IJ}H^{\left(  1\right)
I\rho}\mathcal{F}_{\nu\rho}^{J}+\tfrac{1}{6}Q^{IJ}\left(  \beta C_{ILM}%
H^{\left(  1\right)  L\rho}A^{M\sigma}+H_{I}^{\left(  2\right)  \rho\sigma
}\right)  \mathcal{G}_{J\nu\rho\sigma}\right]  d\Sigma^{\nu}.
\end{equation}
The generalized version of Smarr's formula given in \cite{Gibbons:2013tqa} is
now%
\begin{equation}
M=\tfrac{1}{32\pi G_{5}}\int_{\Sigma_{4}}\left[  H_{I}^{\left(  1\right)
}\wedge\left(  4\mathcal{G}^{I}-\beta C^{IJK}A_{J}\wedge\mathcal{F}%
_{K}\right)  -2H_{I}^{\left(  2\right)  }\wedge\mathcal{F}^{I}\right]  .
\label{Komar mass_5D}%
\end{equation}
Also here note that the differential form,%
\[
4\mathcal{G}^{I}-\beta C^{IJK}A_{J}\wedge\mathcal{F}_{K},
\]
is closed in virtue of $\left(  \ref{dG_5D}\right)  $. As in $\left(
\ref{Komar mass_topological}\right)  $ the explicit Chern-Simons contributions
do only support the purely topological form of the integrand for all values of
$\beta$.

\subsection{Dimensional reduction}

The five-dimensional mass formula in \cite{Gibbons:2013tqa} and the one above
can obviously be reproduced by dimensional reduction of the eleven-dimensional
expression $\left(  \ref{Komar mass_topological}\right)  $.

The five-dimensional fields embed into the eleven-dimensional ones via%
\begin{align}
C  &  =A^{1}\wedge dx^{5}\wedge dx^{6}+A^{2}\wedge dx^{7}\wedge dx^{8}%
+A^{3}\wedge dx^{9}\wedge dx^{10}\label{Field truncations1}\\
F  &  =\mathcal{F}^{1}\wedge dx^{5}\wedge dx^{6}+\mathcal{F}^{2}\wedge
dx^{7}\wedge dx^{8}+\mathcal{F}^{3}\wedge dx^{9}\wedge dx^{10}.
\label{Field truncations2}%
\end{align}

In order to express $G=\star_{11}F$ in terms of the $\mathcal{G}_{I}%
=Q_{IJ}\star_{5}\mathcal{F}^{J}$, we go to a representation in frames:%
\begin{equation}%
\begin{array}
[c]{cccc}%
e^{0}=Z^{-1}\left(  dt+k\right)  & e^{i}=\sqrt{\gamma_{ii}}dx^{i} &
e^{5}=\left(  \tfrac{Z_{2}Z_{3}}{Z_{1}}\right)  ^{\frac{1}{6}}dx^{5} &
e^{6}=\left(  \tfrac{Z_{2}Z_{3}}{Z_{1}}\right)  ^{\frac{1}{6}}dx^{6}\\
e^{7}=\left(  \tfrac{Z_{1}Z_{3}}{Z_{2}}\right)  ^{\frac{1}{6}}dx^{7} &
e^{8}=\left(  \tfrac{Z_{1}Z_{3}}{Z_{2}}\right)  ^{\frac{1}{6}}dx^{8} &
e^{9}=\left(  \tfrac{Z_{1}Z_{2}}{Z_{3}}\right)  ^{\frac{1}{6}}dx^{9} &
e^{10}=\left(  \tfrac{Z_{1}Z_{2}}{Z_{3}}\right)  ^{\frac{1}{6}}dx^{10}%
\end{array}
\label{Frames}%
\end{equation}
We compute explicitely the first term of $\left(  \ref{Field truncations2}%
\right)  $ and then find the other two by analogy. It holds:%
\begin{equation}
\star_{11}\left(  e^{\mu}\wedge e^{\nu}\wedge e^{5}\wedge e^{6}\right)
=\star_{5}\left(  e^{\mu}\wedge e^{\nu}\right)  \wedge e^{7}\wedge...\wedge
e^{10}, \label{frame dual}%
\end{equation}
which can be rewritten with $\left(  \ref{Frames}\right)  $ to%
\begin{equation}
\left(  \tfrac{Z_{2}Z_{3}}{Z_{1}}\right)  ^{\frac{1}{3}}\star_{11}\left(
e^{\mu}\wedge e^{\nu}\wedge dx^{5}\wedge dx^{6}\right)  =\left(  \tfrac
{Z_{1}Z_{3}}{Z_{2}}\right)  ^{\frac{1}{3}}\left(  \tfrac{Z_{1}Z_{2}}{Z_{3}%
}\right)  ^{\frac{1}{3}}\star_{5}\left(  e^{\mu}\wedge e^{\nu}\right)  \wedge
dx^{7}\wedge...\wedge dx^{10}.
\end{equation}
and thus becomes with $\left(  \ref{Reparameterization}\right)  -\left(
\ref{G_5D}\right)  $%
\begin{equation}
\star_{11}\left(  \mathcal{F}^{1}\wedge dx^{5}\wedge dx^{6}\right)  =\tfrac
{1}{X_{1}^{2}}\star_{5}\mathcal{F}^{1}\wedge dx^{7}\wedge...\wedge
dx^{10}=2Q_{11}\star_{5}\mathcal{F}^{1}\wedge dx^{7}\wedge...\wedge
dx^{10}=2\mathcal{G}_{1}\wedge dx^{7}\wedge...\wedge dx^{10}.
\end{equation}
Analogously proceeded for $\mathcal{G}_{2}$ and $\mathcal{G}_{3}$, we finally
achieve%
\begin{equation}
G=2\left(  \mathcal{G}_{1}\wedge dx^{7}\wedge dx^{8}\wedge dx^{9}\wedge
dx^{10}+\mathcal{G}_{2}\wedge dx^{5}\wedge dx^{6}\wedge dx^{9}\wedge
dx^{10}+\mathcal{G}_{3}\wedge dx^{5}\wedge dx^{6}\wedge dx^{7}\wedge
dx^{8}\right)  . \label{Field truncations3}%
\end{equation}

With $\left(  \ref{F_Killing}\right)  $, $\left(  \ref{G_Killing}\right)  $,
$\left(  \ref{F_Killing_5D}\right)  $, $\left(  \ref{G_Killing_5D}\right)  $,
$\left(  \ref{Field truncations1}\right)  $, $\left(  \ref{Field truncations2}%
\right)  $, and $\left(  \ref{Field truncations3}\right)  $ we find the
relations between the non-trivial forms,%
\begin{align}
H^{\left(  3\right)  }  &  =H^{\left(  1\right)  1}\wedge dx^{5}\wedge
dx^{6}+H^{\left(  1\right)  2}\wedge dx^{7}\wedge dx^{8}+H^{\left(  1\right)
3}\wedge dx^{9}\wedge dx^{10}\\
H^{\left(  6\right)  }  &  =2H_{1}^{\left(  2\right)  }\wedge dx^{7}\wedge
dx^{8}\wedge dx^{9}\wedge dx^{10}+2H_{2}^{\left(  2\right)  }\wedge
dx^{5}\wedge dx^{6}\wedge dx^{9}\wedge dx^{10}\nonumber\\
&  +2H_{3}^{\left(  2\right)  }\wedge dx^{5}\wedge dx^{6}\wedge dx^{7}\wedge
dx^{8},
\end{align}
and the remaining forms,%
\begin{align}
\lambda^{\left(  2\right)  }  &  =\lambda^{1}dx^{5}\wedge dx^{6}+\lambda
^{2}dx^{7}\wedge dx^{8}+\lambda^{3}dx^{9}\wedge dx^{10}\\
\Lambda^{\left(  5\right)  }  &  =2\Lambda_{1}^{\left(  1\right)  }\wedge
dx^{7}\wedge dx^{8}\wedge dx^{9}\wedge dx^{10}+2\Lambda_{2}^{\left(  1\right)
}\wedge dx^{5}\wedge dx^{6}\wedge dx^{9}\wedge dx^{10}\nonumber\\
&  +2\Lambda_{3}^{\left(  1\right)  }\wedge dx^{5}\wedge dx^{6}\wedge
dx^{7}\wedge dx^{8}.
\end{align}
If we assume our compact space to be a 6-torus, that is, $M_{6}=T^{6}%
=T^{2}\times T^{2}\times T^{2}$, with $dvol_{6}=dx^{5}\wedge...\wedge dx^{10}%
$, $vol_{6}=\left(  2\pi\right)  ^{6}\overset{6}{\underset{m=1}{\Pi}}r_{m}$,
the $r_{m}$ being the radii of the tori, and use the above stated dictionary,
the eleven-dimensional integral ammounts to%
\begin{align}
&  \int_{\Sigma_{10}}\left[  H^{\left(  3\right)  }\wedge\left(  2G-\alpha
C\wedge F\right)  -H^{\left(  6\right)  }\wedge F\right] \nonumber\\
&  =\int_{\Sigma_{4}\times T^{6}}\left[  H_{I}^{\left(  1\right)  }%
\wedge\left(  4\mathcal{G}^{I}-\alpha C^{IJK}A_{J}\wedge\mathcal{F}%
_{K}\right)  -2H_{I}^{\left(  2\right)  }\wedge\mathcal{F}^{I}\right]  \wedge
dvol_{6}\\
&  =vol_{6}\int_{\Sigma_{4}}\left[  H_{I}^{\left(  1\right)  }\wedge\left(
4\mathcal{G}^{I}-\alpha C^{IJK}A_{J}\wedge\mathcal{F}_{K}\right)
-2H_{I}^{\left(  2\right)  }\wedge\mathcal{F}^{I}\right]  .\nonumber
\end{align}
From this result we see with $\left(  \ref{Komar mass_topological}\right)  $
and $\left(  \ref{Komar mass_5D}\right)  $ that if $\alpha=\beta$ the Komar
masses of both dimensional cases are related like%
\begin{equation}
M_{\left(  11\right)  }=M_{\left(  5\right)  }.
\end{equation}

\section{Conclusion}

We have derived the equations of motion following from eleven-dimensional
supergravity and from that inferred a mass formula of solitonic solutions via
compactifying on a simply connected and Ricci-flat 6-manifold. Furthermore, we
gave the Chern-Simons term an arbitrary constant coefficient in the action to
see in how far the results are influenced by this parameter. We finally
obtained a generalized version of Smarr's formula and also showed how to
arrive back at the five-dimensional theory by performing the
eleven-dimensional Komar integral over $T^{6}$.

The most intriguing aspect of the calculations done both in
\cite{Gibbons:2013tqa} and here is the proof of the possibility of contructing
massive soliton solutions without the need of horizons and so showing that the
techniques used in microstate geometries are the only methods that can support
solitons. Moreover, we could show that making the Chern-Simons term arbitrary
does not change this fact; the incorporation of Chern-Simons interactions does
not yield extra pieces in the mass formula in addition to the topological
terms, but is only significant for the purely topological nature of the
soliton mass.

The question, whether the eleven-dimensional generalization does indeed
contain new physics in its spectrum of different topological mass terms, is
yet to be investigated.\bigskip\medskip

\leftline{\bf Acknowledgements}First and foremost would I like to thank
Nicholas Warner for his helpful and inspiring advising through my time as PhD
student, as well as for his great style of teaching. Moreover I am thankful
for helpful discussions with Benjamin Niehoff and my other fellow researchers
of the High Energy Group of USC. My gratitude also goes to the University of
Southern California for offering all its beautiful places to be and giving the
opportunities to\ study and do work in manifold kinds in a comfortable and
vitalising atmosphere. Last but not least would I like to thank the DOE for
their support under the grant DE-FG03-84ER-40168.

\end{document}